\pdfoutput=1 
\documentclass[journal]{IEEEtran}
\pdfoutput=1
\usepackage{lipsum}
\usepackage[utf8]{inputenc}
\DeclareUnicodeCharacter{2212}{-}
\usepackage{graphicx}
\usepackage{amsmath}
\usepackage{multirow}
\usepackage{verbatimbox}
\usepackage{caption}
\DeclareCaptionLabelSeparator{none}{ }
\usepackage{placeins}
\usepackage{tabularx}
\usepackage{afterpage,lipsum}
\usepackage{float}
\usepackage{dblfloatfix}
\usepackage{algorithm}
\usepackage[noend]{algpseudocode}
\makeatletter
\def\BState{\State\hskip-\ALG@thistlm}
\makeatother

\ifCLASSINFOpdf
\else
\fi
\begin{document}
%
\title{A Generative Machine Learning-Based Approach for Inverse Design of Multilayer Metasurfaces}
%
%
%

\author{Parinaz Naseri and ~Sean~V. Hum \\ submitted to IEEE Transactions on Antennas and Propagation
        
\thanks{P. Naseri and S. V. Hum are with the Edward S. Rogers Sr. Department of Electrical and Computer Engineering, 10 King's College Road, toronto, Ontario, Canada, M5S3G4, email: parinaz.naseri@utoronto.ca }}

\maketitle

\begin{abstract}
The synthesis of a metasurface exhibiting a specific set of desired scattering
properties is a time-consuming and resource-demanding process, which conventionally relies on many cycles of full-wave simulations. It requires an experienced designer to choose the number of the metallic layers, the scatterer shapes and dimensions, and the type and the thickness of the separating substrates. Here, we propose a generative machine learning (ML)-based approach to solve this one-to-many mapping and automate the inverse design of dual- and triple-layer metasurfaces. Using this approach, it is possible to solve multiobjective optimization problems by synthesizing thin structures composed of potentially brand-new scatterer designs, in cases where the inter-layer coupling between the layers is non-negligible and synthesis by traditional methods becomes cumbersome. Various examples to provide specific magnitude and phase responses of $x$- and $y$-polarized scattering coefficients across a frequency range as well as mask-based responses for different metasurface applications are presented to verify the practicality of the proposed method. 
\end{abstract}


%
\IEEEpeerreviewmaketitle

\section{Introduction}
%
%
%
%
\IEEEPARstart{E}{lectromagnetic} metasurfaces (EMMSs) are 2D structures composed of sub-wavelength uniform or non-uniform unit cells composed of metallic scatterers and/or dielectric substrates. They provide the ability to manipulate electromagnetic waves in extraordinary ways such as spectrum filtering, wave manipulation, and polarization conversion \cite{OQT}. The design of an EMMS traditionally includes two steps. The first is to map the tangential field transformations on the two sides of the EMMS to macroscopic properties such as scattering parameters, susceptibilities \cite{Achouri}, or surface impedance/admittance \cite{Epstein}. The second is to implement these properties using a physical unit cell structure. However, there is no straightforward method to solve this step and empirical approaches that involve \emph{ad hoc} design procedures have been mostly employed so far. This approach relies on optimization loops of time-consuming and resource-demanding full-wave simulations. While experience can expedite the design process, adequate exploration of the design space remains challenging.

\begin{figure}[!ht]
	\centering 
  	\includegraphics[width=3.3in]{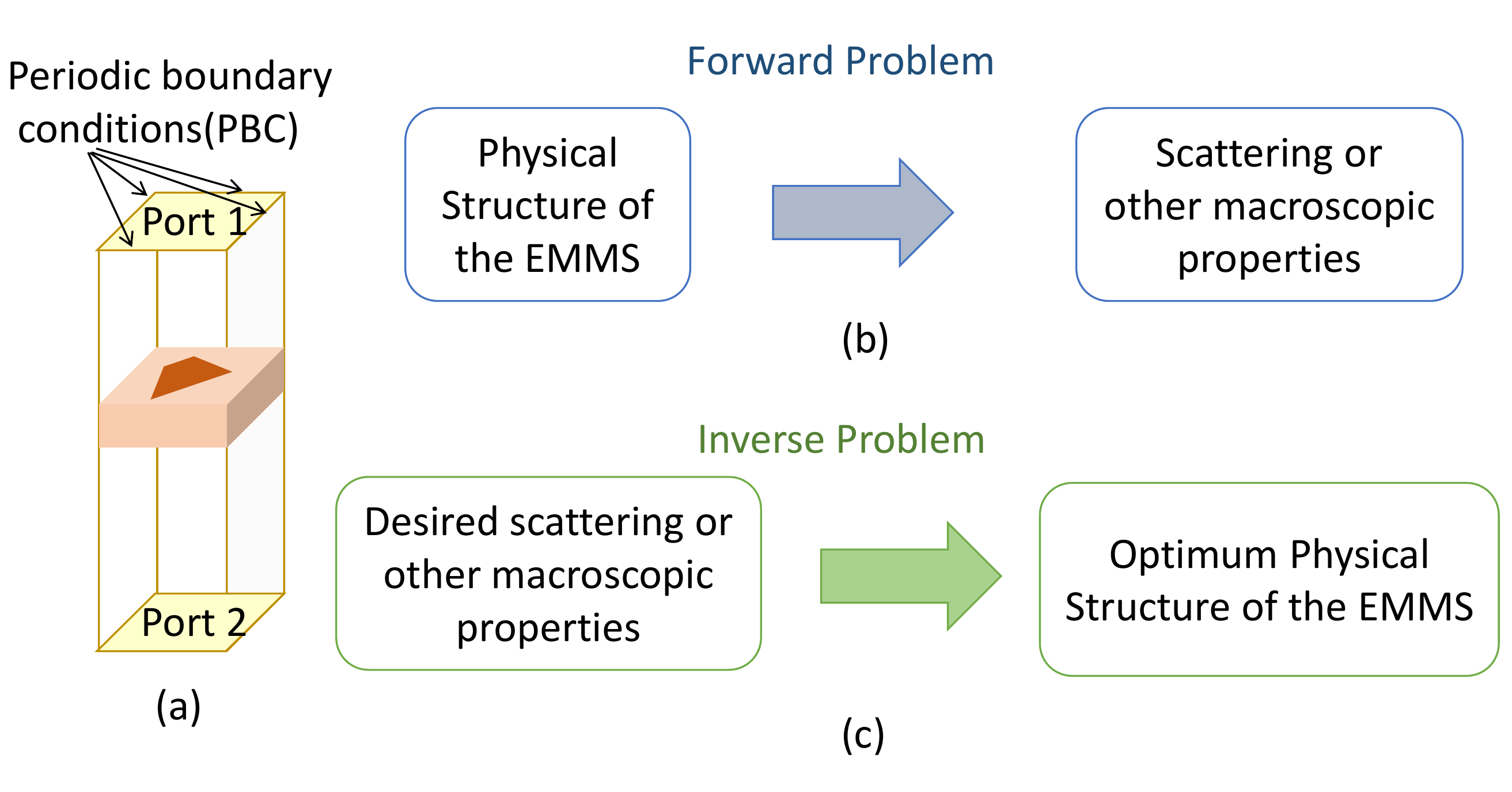}
  	\caption{(a) Simulation setup to obtain the scattering coefficients of a metasurface unit cell, (b) forward, and (b) inverse problem of EMMS design.}
  	\label{Fig1aLabel}
\end{figure}

Scattering properties of EMMS unit cells are usually evaluated when they are surrounded by similar unit cells in the same plane and excited by transverse electric (TE) or transverse magnetic (TM) waves from both sides. The simulation setup is implemented by stipulating periodic boundary conditions on the perpendicular sides of the plane and two excitation ports on the top and bottom of the unit cell, shown in Fig. \ref{Fig1aLabel} (a). Deep learning models based on regression can solve the forward problem of predicting the scattering properties given the physical parameters of the EMMS, shown in Fig. \ref{Fig1aLabel} (b). Such prediction tools can be used as surrogate models in analysis and optimization of non-uniform metasurfaces \cite{arrebola}--\cite{b7}. However, the inverse problem of predicting the physical structure of the EMMS based on the desired properties, shown in Fig \ref{Fig1Label} (c), is not easy to solve. The reason is that this \emph{inverse problem} is a one-to-many mapping, meaning that one set of desired scattering properties might be potentially provided by many different EMMSs. By training potentially many sub-models for different regions of the solution space, it is possible to find the optimum design for simple structures \cite{b8}--\cite{b9}. In the general problem where the scatterer on the different layers of the EMMS can have various types of shapes, this  approach can be inefficient and complicated. More importantly, these models lack the ability to explore the broader design space and generate new plausible solutions if required.
    
Generative machine learning (ML) techniques have the capability to effectively capture the underlying patterns in large complex datasets and employ that knowledge to generate new structures \cite{material}--\cite{Seq}. Several machine learning (ML) techniques for image processing have been used for the synthesis of single-layer microwave \cite{b10}--\cite{b11} and dielectric-based optical \cite{b12}--\cite{b14} EMMSs. Among the generative models, generative adversarial networks (GANs) \cite{b11}--\cite{b14} are able to produce similar examples to the ones in the training dataset by developing an appropriate generator and critic \cite{b15}. For generating an EMMS with desired scattering coefficients using GANs, a training set with the same or similar properties is required. Obtaining such a dataset is more expensive and time-consuming than solving the inverse design problem itself.

Another type of generative model known as variational autoencoders (VAE) \cite{Kingma} consists of an encoder and a decoder. A VAE can convert high-dimensional discrete data to a low-dimensional continuous space known as the latent space, where standard optimization techniques can be applied to finding solutions \cite{b18}, which is not easily applicable in the initial form of the input. By training the VAE, the model learns the conditional probability distribution of the input given the latent variables. Therefore, it is possible to generate new plausible inputs by drawing samples from the latent space and feeding them to the learned distribution. The VAE has been previously employed to represent single- and dual-layer metasurfaces with negligible interlayer coupling using a continuous latent space \cite{b19}. However, the optimum latent variable is found through \emph{ad hoc} random selection instead of an optimization technique. Hence, this approach may miss an opportunity to exploit one of the important advantages of the VAE to accelerate optimization.

Due to their increased degrees of freedom,  multilayer EMMSs are both promising solutions and difficult to optimize for multiobjective electromagnetic problems \cite{george}--\cite{Zhang}. Placing the layers closely using thin substrates offers additional degrees of freedom in terms of the order of behavior, but at the same time, it complicates the design further. Conventionally, designing and optimizing structures with high interlayer coupling is difficult. It either requires sophisticated equivalent circuit models (ECMs) (and corresponding simulations to find the values of the ECMs) \cite{b20}--\cite{b21}, or brute-forcing the problem. In general, the solution space to design a multilayer EMMS includes different categories of scatterer shapes. This space is potentially high-dimensional where it is difficult to find the global optimum, and yet might not include the optimum design due to its limited domain. Therefore, an automated generative tool is required that first traverses the design space efficiently to find the global optimum solution within it and also expands the solution space if it is inadequate. Such a tool to design an EMMS with the desired scattering properties could greatly aid in designing surfaces with next-generation capabilities.
   
Here, we propose a machine learning-based approach to solve the \emph{inverse problem} by predicting the configuration of thin multilayer EMMS structures given the desired scattering properties of the surface. We leverage the interlayer coupling as an extra degree of freedom to match the desired properties with the minimum error. For that, the ML tool is trained by thin EMMSs composed of known shapes of patch- and slot-based scatterers. Using canonical structures instead of pixelated EMMSs, we can transfer our intelligence and experience to ML models with less training data to bring to bear the fact that a great deal of fundamental EMMS behavior can be synthesized using these structures that do not require dense description. This tool exploits the knowledge obtained from the training data and explores the solution space by generating new structures according to the new behavior they offer from the canonical shapes.  

This approach utilizes VAE and regression ML models as the scattering property predictors to generate multilayer EMMSs based on the desired scattering coefficients. We use the proposed scheme to represent the sample EMMSs in the training database with a continuous low-dimensional latent space where interpolation, exploration, and optimization can be easily performed. Each latent variable represents not only the physical parameters of the EMMS, but also its scattering properties. Regularizing the latent space in this way greatly impacts the success of the optimization algorithm to  efficiently converge to the global optimum. We evaluate the objective through a hybrid approach of invoking computation simulations when needed, and surrogate models when they are known to provide sufficient accuracy, to accurately guide the design process.

This paper is organized as follows. The variational autoencoder is described as a generative model and compared to conventional autoencoder in Section \ref{section:II}. The proposed approach to represent multilayer metasurfaces with a low-dimensional and continuous latent space, where optimization is performed to find the optimum structure, is explained in Section \ref{section:III}. The details about the training data of the multilayer metasurfaces are presented in Section \ref{section:IV}. The potential of the proposed approach is demonstrated through various examples in Section \ref{section:V}. Full generative capability of the the proposed approach is shown in Section \ref{section:VI}. The conclusions are drawn in Section \ref{section:Conclusion}. The architecture of the neural networks are detailed in Appendix \ref{appendix: B}. 

\section{The Variational Autoencoder: a Generative Model}\label{section:II}

\begin{figure}[!h]
	\centering 
  	\includegraphics[width=3.3in]{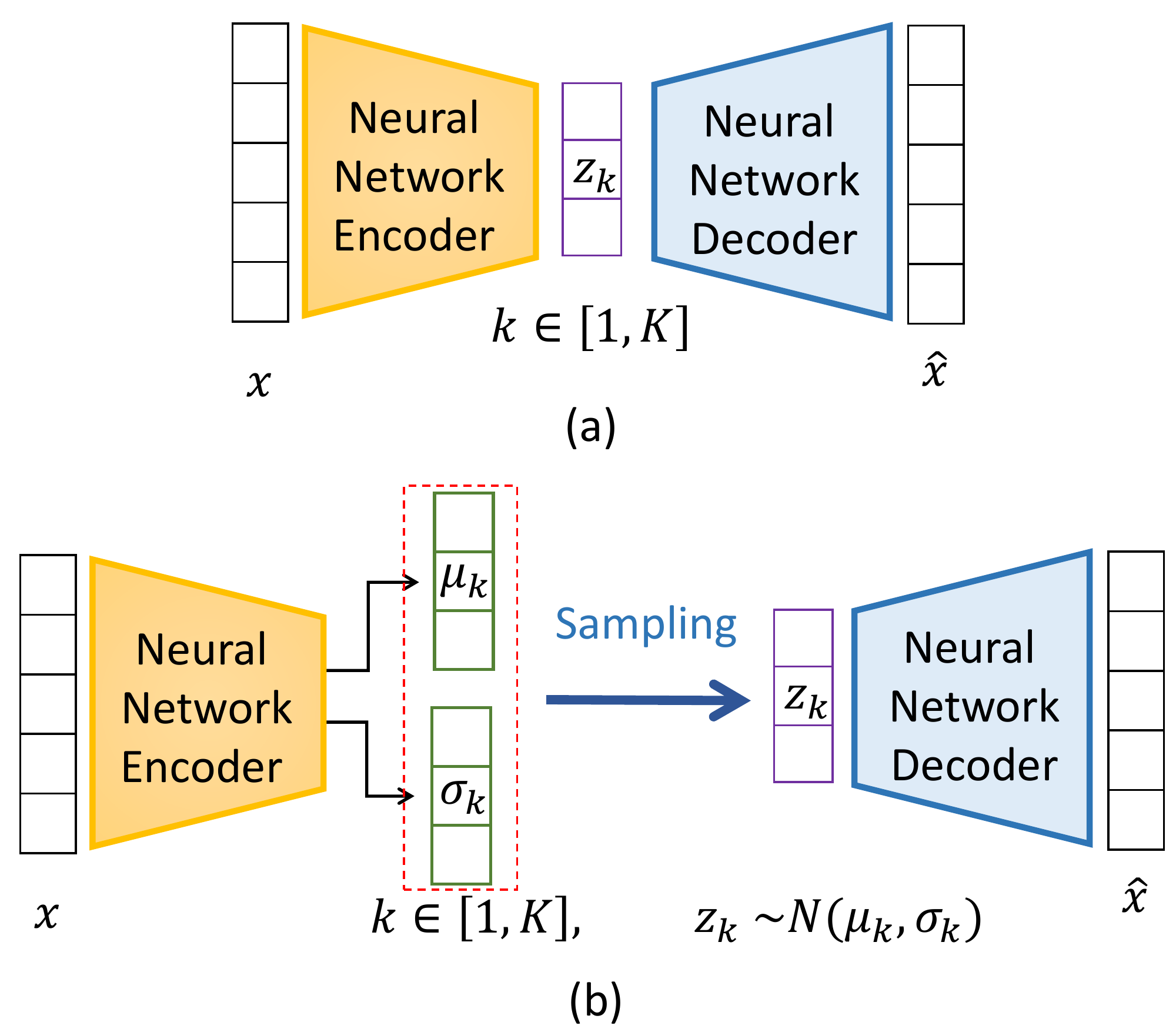}
  	\caption{(a) Conventional autoencoder and (b) variational autoencoder (VAE).}
  	\label{Fig1Label}
\end{figure}

A conventional autoencoder (AE) is an encoder followed by a decoder, shown in Fig. \ref{Fig1Label} (a). The encoder is a neural network whose input and output are a data point \emph{x} and a hidden representation \emph{z}, respectively. Each $x$ is described by $N$ features and is $N$-dimensional; each $z$ is described by $K$ number of features and is $K$-dimensional, where $K<N$. The decoder is another neural network whose input and output are the latent representation $z$ and the reconstructed $x$, respectively. The general idea of the AE is to learn the best encoding-decoding scheme using an iterative optimization process. The neural networks architectures are optimized to minimize the difference between the original $x$ and the encoded-decoded value $\hat{x}$, i.e. reconstructed $x$. Thus, the overall AE architecture creates a ``bottleneck'' for data $x$ that ensures only the main structured part of the information can go through by the encoder and be reconstructed by the decoder. AEs are employed to represent the data with a lower dimensional space $z$ and remove noise in it. Moreover, samples with similar features will be clustered together as different separated regions in the latent space $z$.

A variational autoencoder (VAE) is similarly the combination of an encoder and a decoder, shown in Fig. \ref{Fig1Label} (b). However, the goal of the VAE is to learn  a probability distribution $P(x)$ over a multidimensional variable $x$. By modelling the distribution, it is possible to draw samples from the distribution to create new plausible values of $x$. Fig. \ref{Fig1Label} (b) shows the scheme of a VAE. In practice, the encoded distributions of the VAE are chosen to be normal so that the encoder can be trained to return the mean, $\mu_{k}$, and the variance, $\sigma_{k}$, vectors that describe these Gaussians, where  $k\in[1,K]$. The latent space, $z$, is created by drawing samples from these distributions. By learning the distribution of $z$ described by  $\mu_{k}$ and $\sigma_{k}$ instead of $z$ itself, the VAE can be used as a generative model unlike the conventional autoencoders \cite{b15}.

Once the VAE is trained, its decoder specifically can be used as a \emph{generative model} that outputs parameters of the likelihood distribution of $p(x|z)$. It means that during the training process, the decoder learns to reconstruct the data $x$ given a representation $z$. After the training, it can generate new examples of $x$ given new samples from the latent space based on their difference to the training samples. Depending on initial distribution of data in $x$, if the dimension of the latent space, $K$, is set to be too small, important information in $x$ will be lost when converted to $z$ by the encoder. Therefore, the decoder cannot fully reconstruct $x$ from $z$. On the other hand, if $K$ is chosen to be too large, it defeats the purpose of the VAE and the latent space becomes sparse.   

The loss function of a VAE, denoted by $L_{VAE}$, is composed of two terms for each data point $x$, \begin{equation}\label{eq:1}
L_{VAE} = L_{recons} + L_{KL}.
\end{equation} 
This loss function is minimized over all possible $x$ in the training data set to optimize the architecture of the VAE. The first part is the ``reconstruction loss'', 
\begin{equation}\label{eq:1a}
L_{recons} = |x-\hat{x}|^2. 
\end{equation} 
Minimizing $L_{recons}$ helps to convert $x$ to $z$ and reconstruct $\hat{x}$ from $z$, where $x$ and $\hat{x}$ are ideally equal. Therefore, the VAE learns to describe $x$ with $K$ variables instead of $N$ variables, where $K < N$. The second term is the ``regularization term'' that tends to organize the latent space in such a way that the distributions returned by the encoder are close to the normal distribution, $N(0,1)$. The difference between the latent space distribution and the standard normal distribution can be expressed by the Kulback-Leibler (KL) divergence \cite{b22}, denoted by $L_{KL}$ with the closed form 
\begin{equation}\label{eq:2}
	\begin{array}{l}
 L_{KL} = \sum_{k=1}^{K} KL[N(\mu_k,\sigma_k), N(0,1)] \\
 \quad \quad \: =\frac{1}{2} \sum_{k=1}^{K} (\sigma_{k}+ \mu_{k}^2 - \log(\sigma_{k})-1).
 \end{array}
\end{equation} 
This term gives the VAE its main and distinct feature compared to conventional AEs, which is encoding a set of datapoints to a continuous and differentiable latent space. It means that the VAE's latent space does not have large gaps that would exist in the latent space of the conventional autoencoder. Over such a space, different optimization techniques such as gradient-descent \cite{b18} can be performed efficiently.

\section{Proposed Approach to Optimize Metasurfaces } \label{section:III}

In the synthesis of a metasurface, mere generation of different scatterer shapes is not enough and creating samples with a specific set of properties is required. If one is to convert the structures of a set of EMMSs to a latent space, $z_{geometry}$, by the scheme shown in Fig. \ref{Fig1Label}, the latent variables will only correspond to different shapes of the scatterers and not their scattering properties. Therefore, optimization over $z_{geometry}$ can be problematic since it results in minimizing an arbitrarily shaped loss function. For example, a loss function to optimize an EMMS,
\begin{equation}\label{eq:2a}
 L_{EMMS} =\frac{1}{n} \sum_{i=1}^{n}|[S]_{target}(i)-[S]_{EMMS}(i)|^2,
\end{equation}
is the mean squared error difference between the scattering properties corresponding to each $z_{geometry}$ and the desired scattering properties, and different from the one in \eqref{eq:1}. The scattering tensor including all the scattering coefficients of each EMMS is denoted with $[S]$ and $n$ is the number of the frequency points.

\begin{figure}[!ht]
	\centering 
  	\includegraphics[width=3.3in]{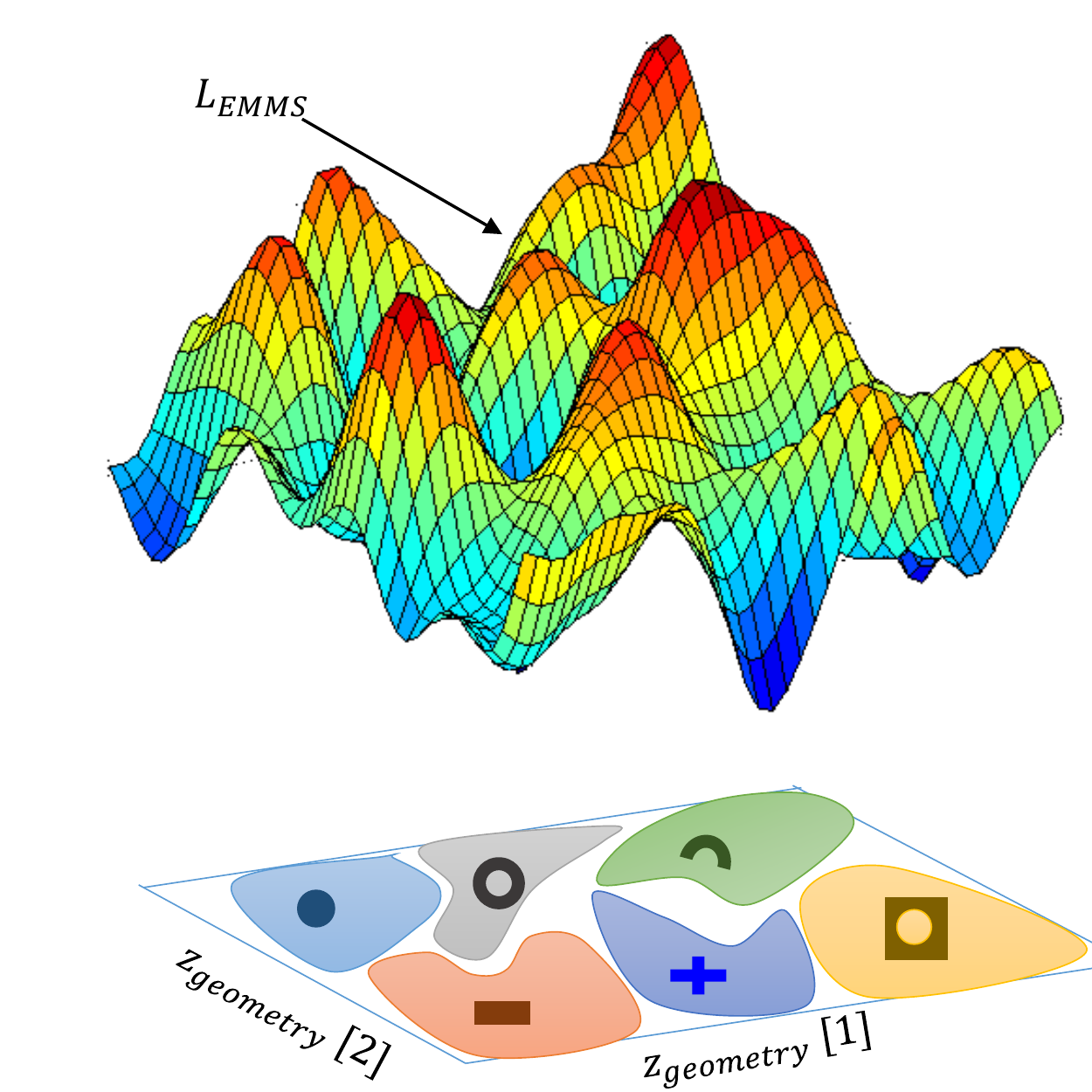}
  	\caption{An arbitrarily shaped EMMS loss function for optimizing a single-layer EMMS to obtain desired scattering coefficients over $z_{geometry}$, obtained by the VAE in Fig. \ref{Fig1Label}.}
  	\label{Fig2Label}
\end{figure}

Let us first consider designing a single-layer EMMS ro realize a specific set of desired scattering coefficients, $[S]_{target}$. The VAE in Fig. \ref{Fig1Label} is used to convert different shapes of scatterers to a 2-dimensional continuous latent space, $z_{geometry}$, where $K=2$ is chosen arbitrarily for this example. The dimension of the latent space is arbitrarily set to be two for ease of visualization in our example. This 2-dimensional latent space, described by $z_{geometry}[1]$ and $z_{geometry}[2]$, is composed of latent variables correspond to different shapes of scatterers on the bottom of Fig. \ref{Fig2Label}. The corresponding EMMS loss function across the latent space for different scatterer shapes is shown on the top part of in Fig. \ref{Fig2Label}. As it can be see from this figure, the EMMS loss function can be arbitrarily shaped with many local minima. This makes it difficult for optimization methods to converge to a global optimum. There are also some gaps in the latent space not covered by the samples in the training set, where the optimum latent variable might lie. Therefore, we first regularize a latent space according to both the shape and scattering properties of the EMMSs. Furthermore, we employ a generative model to produce new structures based on the target scattering coefficients in case exploring the gaps in the latent space becomes advantageous. 

\subsection{Using the Latent Space to Represent Metasurfaces}

We adapt the scheme of the regular VAE shown in Fig. \ref{Fig1Label} so that the variables in the latent space not only represent different structures of the EMMSs but also the scattering properties they provide. To do so, we employ an extra neural network model as a predictor alongside the VAE. The predictor solves the \emph{forward problem} shown in Fig. \ref{Fig1aLabel} (a) so it outputs the scattering coefficients, however, its input is the latent variable instead of the EMMS physical parameters. Since the latent variables are stochastic, we use the mean value of the latent variable, $\mu_{k}$, as the input of the predictor for better training. We also use the frequency points as an extra input to the predictor to include the dispersive behavior of the EMMS. 

\begin{figure}[!ht]
	\centering 
  	\includegraphics[width=3.5in]{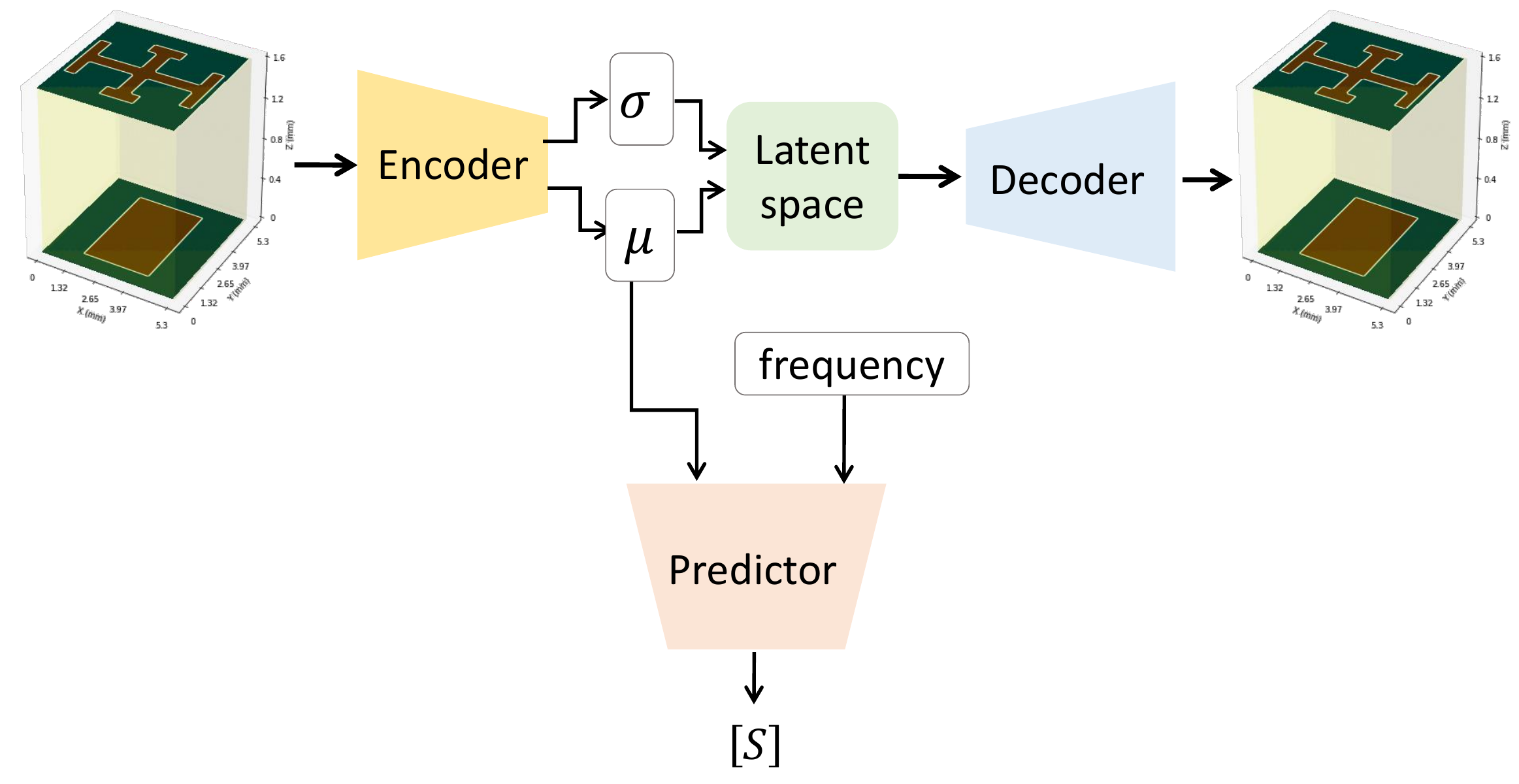}
  	\caption{The proposed approach to regularize a latent space where both the the physical shape of the metasurfaces and their scattering properties change smoothly for better optimization.}
  	\label{Fig3Label}
\end{figure}

Fig. \ref{Fig3Label} shows the scheme of the proposed approach. By jointly training the VAE and the predictor, we can organize the latent space such that close variables in this space not only have similar physical shapes, but they also present similar scattering coefficients. This continuous latent space can be used for meaningful interpolation, efficient optimization, and adequate exploration. Moreover, the loss function for optimizing the EMMS, $L_{EMMS}$, gets shaped in a way that the global optimum can be easily found. 
\begin{figure*}[ht!]
	\centering 
  	\includegraphics[width=5.5in]{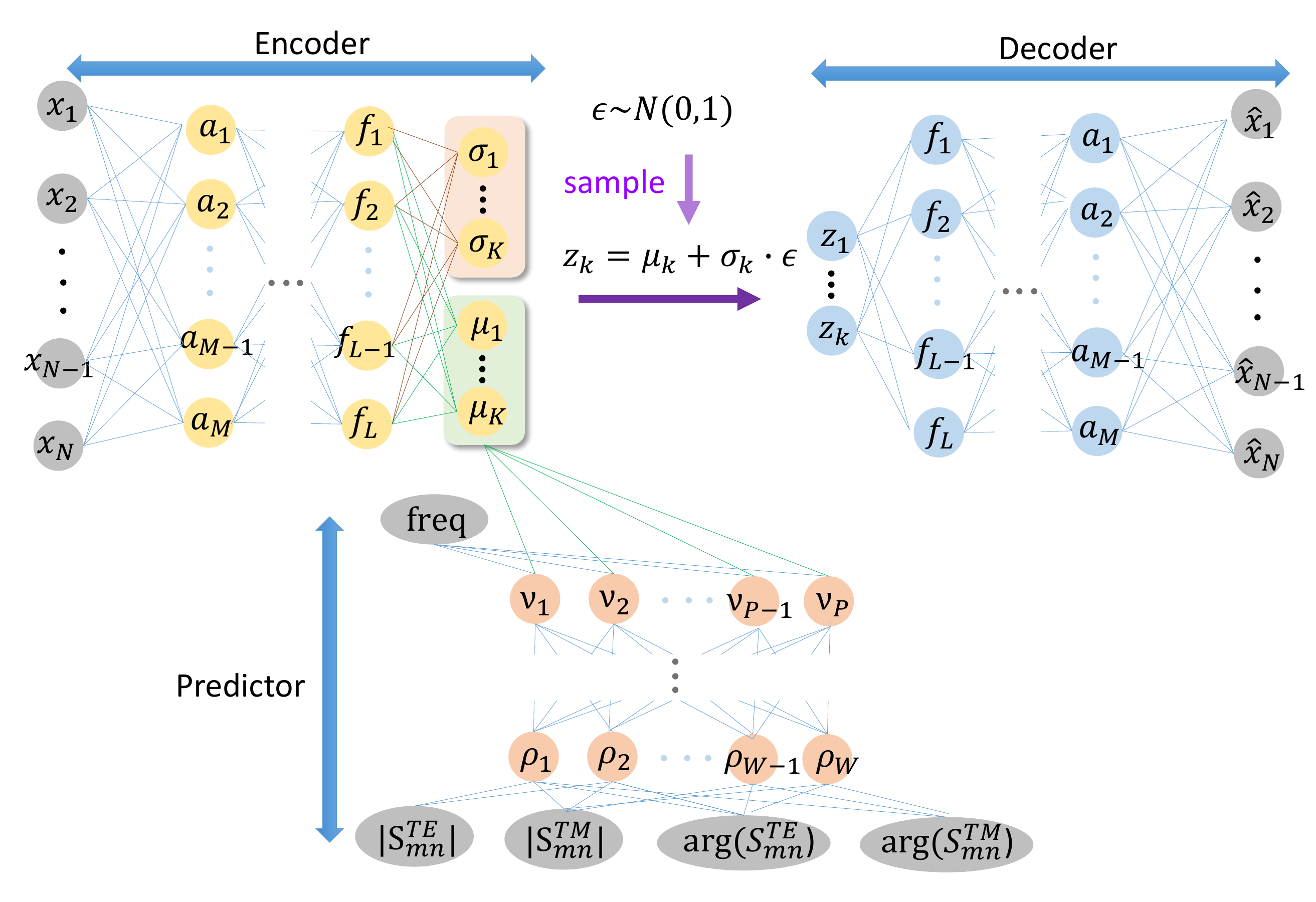}
  	\caption{Implementation of the VAE plus predictor using fully-connected multilayer perceptrons (MLP), where $m,n \in{\{1,2\}}$.}
  	\label{Fig4Label}
\end{figure*}

\subsection{Training the Proposed ML Models}
The scheme shown in Fig. \ref{Fig3Label} is implemented using sets of fully-connected multilayer perceptrons shown in Fig. \ref{Fig4Label}. The loss function, $L_{VAEpred}$, is used to jointly train the neural networks shown in Fig. \ref{Fig4Label}, and is defined as
\begin{equation}\label{eq:3}
	\begin{array}{l}
L_{VAEpred} = \alpha \times L_{recons} + L_{KL} + \beta \times L_{pred}. 
	\end{array}
\end{equation}
$L_{VAEpred}$ is the weighted sum of the reconstruction loss, the KL divergence \eqref{eq:2}, and the prediction loss. $\alpha$ and $\beta$ are weights and hence hyperparameters to be tuned to achieve satisfactory results in the reconstruction and the prediction processes. The reconstruction loss, $L_{recons}$, depends on the type of data in the input $x$. If $x$ consists of continuous variables, the reconstruction loss can be expressed using \eqref{eq:1a}. Otherwise, if the variables composing the input have discrete values such as $0$ and $1$ as in the case of black and white images, we denote it with $X$. The difference between $X$ and $\hat{X}$, reconstructed input, can be expressed as the cross-entropy loss,
\begin{equation}\label{eq:4}
 L_{recons}= −\sum_{j=1}^{N}(X_j \log(\hat{X_j})+(1−X_j) \log(1−\hat{X_j})). 
\end{equation}
Here, each EMMS is described by the images of its layers. Therefore, $x$ consists of $0$ and $1$ for non-metallic and metallic parts of each layer, respectively. Hence, the reconstruction loss is calculated using \eqref{eq:4}. 

The scattering properties of the multilayer EMMSs are obtained through a process of cascading the general scattering matrices (GSMs) describing its constituent scatterers \cite{b24}. The GSMs include high-order scattering coefficients alongside the fundamental ones. Therefore, by doing so, we are able to capture the response of the EMMS including its interlayer coupling. The cascading process is fast and inexpensive. Therefore, creating the training data in this way is both very efficient and resource-saving since in case a new EMMS is composed of a scatterer, the scatterers's GSM can be reused. The prediction loss, $L_{pred}$, in \eqref{eq:3} can be calculated based on the difference between the actual scattering coefficients of the EMMSs in the training set and the predicted ones using different loss functions such as the mean squared error.    

The weights in the neural networks shown in Fig. \ref{Fig4Label} are optimized using the backpropagation algorithm \cite{b23} by computing the derivatives of the loss function \eqref{eq:3} with respect to each layer's weights. As mentioned earlier, the latent space $z$ is stochastic and formed by sampling from a normal distribution described by $\mu_k$ and $\sigma_k$. However, it is not possible to differentiate through this step and update the neural networks parameters before it in the backpropagation algorithm. Therefore, a reparameterization trick \cite{Kingma} can be performed to sample a variable $\epsilon$ with standard distribution and draw samples from the intended Gaussian as in
\begin{equation}\label{eq:5}
	\begin{array}{l}
z_k = \mu_k + \sigma_k \cdot \epsilon_k, \quad  k \in [1,K] \quad \textrm{and} \quad
 \epsilon \sim N(0,1). 
	\end{array}
\end{equation} 
Since there is no need for the backpropagation algorithm to pass down this sampling branch, the derivatives can be computed as usual. The neural networks are implemented and trained using the TensorFlow backend Keras in Python. 

\subsection{Optimum Metasurface}

\begin{figure}[tp!]
	\centering 
  	\includegraphics[width=3.5in]{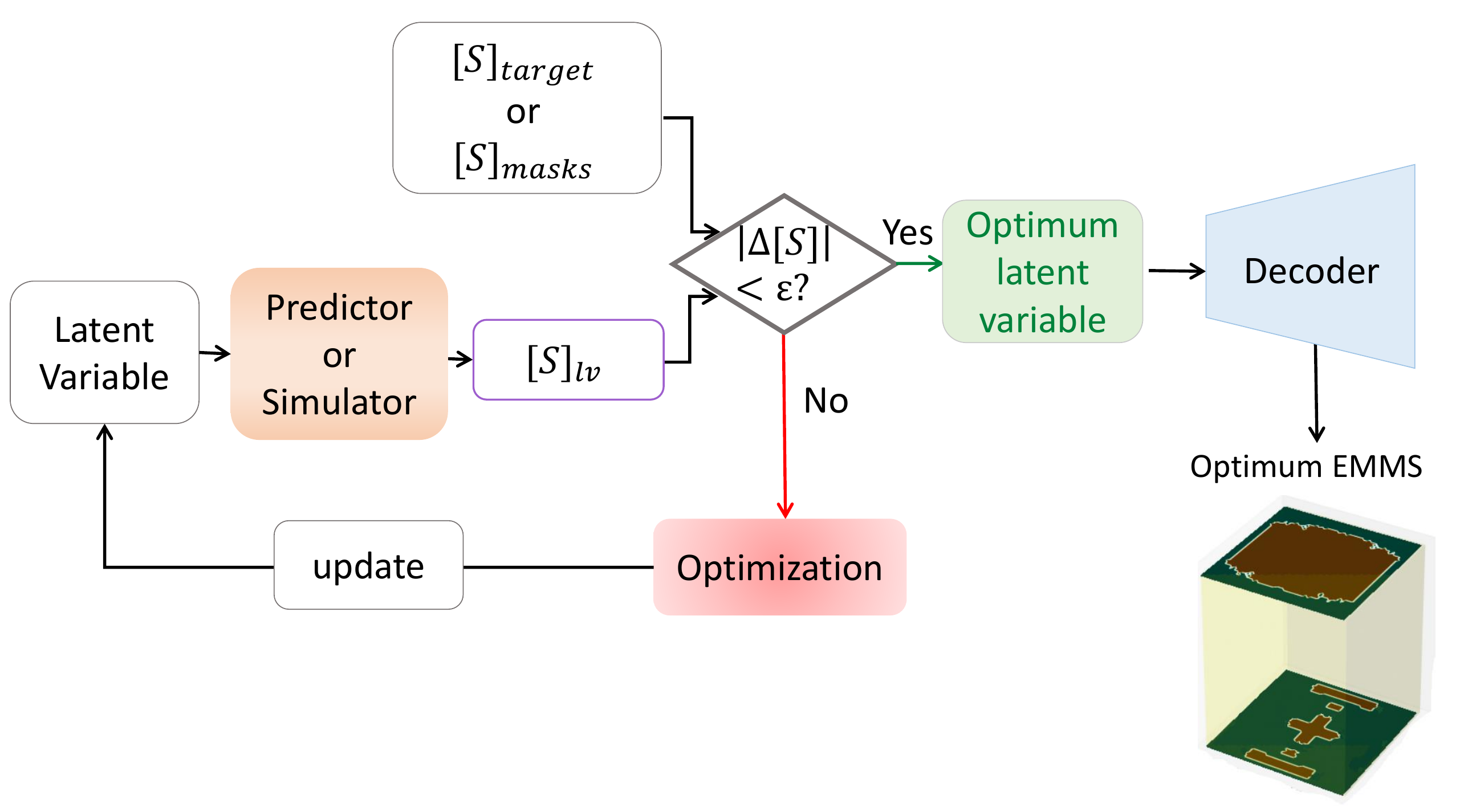}
  	\caption{Optimization in the latent space to obtain the optimum latent variable and converting it to the optimum EMMS.}
  	\label{Fig5Label}
\end{figure}

Once the VAE and the predictor in Fig. \ref{Fig3Label} are trained, the latent variables will be compact representations of the metasurfaces, representing both their physical structures and scattering coefficients. Therefore, based on the target scattering coefficients, $[S]_{target}$, any standard optimization algorithm can be employed to find the optimum latent variable. Then, the optimum latent variable is input to the decoder and the physical implementation of the optimum EMMS is obtained. These steps are illustrated in Fig. \ref{Fig5Label}. 

To obtain the scattering properties corresponding to each latent variable, $[S]_ \textrm{lv}$, it matters whether the latent variable is the compact representation of a known EMMS, i.e. it exists in the training data, or not. If the latent variable is not close to any of the latent variables of the training EMMSs, that means it represents a new EMMS. For such cases, one cannot rely on the predictor to obtain the accurate results for the scattering coefficients of the EMMS. This is mainly because for thin EMMSs where inter-layer coupling is significant, slight changes in the scatterers' shapes and dimensions can result in significantly different results. The ML models are not capable capturing such relations unless they have access to a significant amount of training data, which is not efficient and desirable for EM applications. To address this issue, we use a three-case method to obtain $[S]_ \textrm{lv}$ based on the similarity between the EMMS decoded from the latent variable, $X_ \textrm{lv}$, and the EMMSs in the training data, $X_t$. 

If the latent variable gets decoded to an EMMS in the training data or a close structure in that set, the ML predictor is used to output $[S]_ \textrm{lv}$. If the minimum mean squared difference between the decoded EMMS and the EMMSs in the training data is less than some threshold, $\min(\sum_{j=1}^{N}|X_ \textrm{lv}-X_t|)<\gamma$, the predictor is used. $\gamma$ is specified in Section \ref{section:IV} and is based on the primitives used in the training data. A primitive is a scatterer shape corresponding to one of the canonical ones we specify. However, if the latent variable gets decoded to a new EMMS, a simulation-based approach is used. This new EMMS might be a new combination of the known scatterer shapes, explained in Section \ref{section:IV}, or combination of new generated scatterer shapes. In the former case, we use a fast inexpensive process to cascade the stored GSMs of the known scatterers and obtain the scattering properties of the EMMS. In case the new EMMS is composed of a new scatterer shape, the image of the scatterer is meshed using the Rao-Wilton-Glisson (RWG) basis functions and fed to an in-house spectral-domain periodic method of moments (MoM)-based simulation tool employing those basis functions. The GSM of this scatterer is calculated and saved. This process is done for all the layers of the EMMS. The GSMs of these scatterers are similarly cascaded to obtain $[S]_ \textrm{lv}$. Once $[S]_ \textrm{lv}$ is obtained using any of the three aforementioned ways, $|\Delta[S]|$ can be computed analogous to \eqref{eq:2a} as
\begin{equation} \label{eq:6}
	|\Delta[S]| =\frac{1}{n} \sum_{i=1}^{n} |[S]_{target}(i) -[S]_ \textrm{lv}(i)|^2.  
\end{equation}
It should be noted that using the fast cascading process is a small price to pay to keep the training data small. Otherwise, to make the predictor a reliable simulator, one needs a very large amount of training data preparing the predictor for any change in the shape and dimension of the EMMS scatterers. Moreover, since scatterers with new shapes can have significantly different scattering properties, their full-wave evaluation is necessary to avoid proposing a wrong optimum design.

If $|\Delta[S]|$ is less than some specified criterion, e.g. $\epsilon$, the latent variable under test is considered the optimum solution in the latent space. This optimum latent variable is then fed to the decoder so that the optimum EMMS structure is obtained. To find the global optimum latent variable in the latent space, we use the particle swarm optimization \cite{PSO}. The optimization algorithm is implemented in Python using PySwarms toolkit \cite{pyswarms}.
\vspace{-3mm} 
\section{Dual-Layer and Symmetric Three-Layer EMMS Samples} \label{section:IV}
We apply the proposed approach independently to dual-layer and three-layer EMMSs. In case of three-layer EMMSs, we consider the specific case where the EMMSs have identical top and bottom layers stacked with Rogers $5880$ dielectric slabs ($\epsilon_r = 2.2,\: \mathrm{tan}\delta = 0.0009$). Among the scattering properties, without loss of generality, we train the VAE and the predictor on the labeled data of the TE- and TM- transmission coefficients of the EMMSs. Therefore, EMMSs are optimized to provide desired transmission coefficients. However, one can train the models using the other scattering coefficients (such as reflection coefficients), or even entire sets or subsets of scattering parameters, and optimize EMMSs based on them. We prepare two sets of training data: one for the dual-layer EMMSs and one for the three-layer EMMSs.  

\begin{figure}[!ht]
	\centering 
  	\includegraphics[width=3.5in]{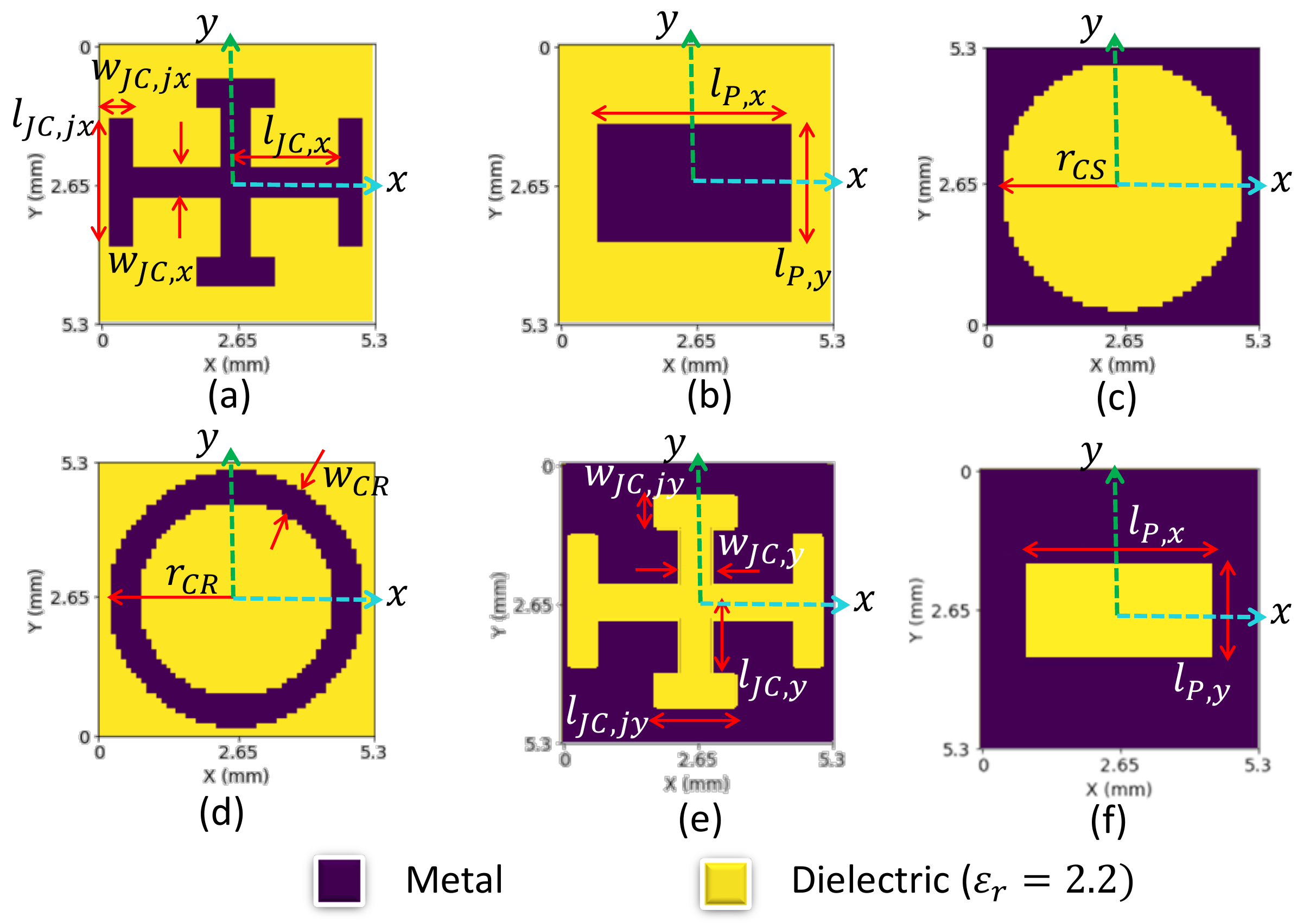}
  	\caption{Primitives used for training: (a) Jerusalem cross (JC), (b) rectangular patch (RP), (c) circular slot (CS), (d) complete ring (CR), (e) complementary Jerusalem cross (compJC), and (f) complementary rectangular patch (compRP).}
  	\label{Fig6Label}
\end{figure}

The primitives shown in Fig. \ref{Fig6Label} with different indicated dimensions in Table \ref{tab:table1} are used for creating EMMSs for training. Based on the range of each parameter, the number of shapes for each shape category is also listed in Table \ref{tab:table1}. We use different shapes of asymmetric scatterers along $x$- and $y$-directions such as a Jerusalem cross (Fig. \ref{Fig6Label} (a)), rectangular patch (Fig. \ref{Fig6Label} (b)), complementary Jerusalem cross (Fig. \ref{Fig6Label} (e)), and complementary rectangular patch (Fig. \ref{Fig6Label} (f)). In addition, some symmetric shapes such as circular slot (Fig. \ref{Fig6Label} (c)) and complete ring (Fig. \ref{Fig6Label} (d)) are used as well. These shapes are chosen based on experience to provide a wide variety of scaterring properties.

Shapes in Fig. \ref{Fig6Label} (a)-(d) are used for the resonator at the air-dielectric interface. Often alternating inductive and capacitive behavior is required in EMMSs with odd number of layers. Therefore, we extend the shape of this scatterer to include complementary Jerusalem cross and complementary rectangular patch in Fig. \ref{Fig6Label} (e)-(f). The scatterers in dual- and three-layer EMMSs are simulated from $15$ GHz to $31$ GHz with unit cell period of $5.3$ mm. The periodic boundary conditions on $x$- and $y$-sides are stipulated. The higher-order mode scattering coefficients are the key to capture the inter-layer coupling between the resonators on different layers. Therefore, the excitation is set to be the fundamental and five higher-order modes of $x$- and $y$- directed incident waves. This number of higher-order modes provides sufficiently accurate results to capture the interlayer coupling for the specified frequency range and unit cell period. Each scatterer is translated to meshes using RWG basis functions and fed to our in-house MoM-based simulation tool. The general scattering matrix (GSM) \cite{b24} including the higher-order modes of each simulation is calculated and saved. 

\FloatBarrier 
\captionsetup{skip=2pt}
\vspace{2mm}
\begin{table}[H]
\caption{Dimensions of the Primitives in Fig. \ref{Fig6Label}.}
\centering
\begin{tabularx}{\columnwidth}{c|c|c|c}
    \hline
     \textbf{Shape Type} &  \textbf{Parameter} &  \textbf{Value (mm)} &  \textbf{Num. of Shapes} \\
    \hline \hline
    \multirow{4}{*}{JC $\&$ compJC} & $l_{JC,x/y}$ & $[2.2:0.2:4.0]$ & \multirow{4}{*}{100} \\
    \cline{2-3}
    & $l_{JC,jx/y}$ & $l_{JC,x/y}-1.6$ mm &\\
    \cline{2-3}
    & $w_{JC,x/y}$ & 0.4  & \\
    \cline{2-3}
    & $w_{JC,jx/jy}$ & 0.45 & \\
    \hline
    RP $\&$ comRP & $l_{P,x/y}$ & $[2.0:0.2:5.0]$ & 256 \\
    \hline
    \text{CS} & $r_{CS}$ & $[1.0:0.1:2.6]$ & 17 \\
    \hline
	\multirow{2}{*}{CR} & $r_{CR}$ & $[1.4:0.2:2.6]$ & \multirow{2}{*}{49} \\
    \cline{2-3}
    & $w_{CR}$ & $[0.1:0.2:1.3]$ & \\ 
    \hline
\end{tabularx}
\label{tab:table1}
\end{table}

Dual-layer EMMS training samples are generated by randomly selecting two GSMs and cascading them with different dielectric thicknesses \cite{b24}. About $10,500$ samples with $0.787$ mm-thick dielectric slab and $7000$ samples with $1.575$ mm-thick dielectric slab are used for training. This step is faster and less expensive than the simulation of a dual-layer EMMS. Training samples of the symmetric three-layer EMMSs are generated in a similar way. About $10,000$ three-layer samples with dielectric thickness of $0.787$ mm and $6,500$ samples with $1.575$ mm-dielectric are created. 

For each dual-layer and symmetric three-layer EMMS sample, $52 \times 52$ image-based representations of the two scatterers and the TE- and TM-mode transmission coefficients at each frequency are used as the inputs to train the VAE plus the predictor. We train two separate sets of VAE plus predictor for dual- and three-layer metasurfaces and obtain distinct latent spaces for them. Moreover, since the standard thickness of the dielectric is a discrete value and can change the transmission coefficients drastically, we train different models per thickness as well. Therefore, four different latent spaces to represent dual-layer EMMSs with thicknesses of $0.787$ mm and $1.575$ mm and three-layer EMMSs with total thickness of $1.574$ mm and $3.015$ mm are obtained. Information about the neural network architectures and their training process is detailed in Appendix \ref{appendix: B}. Based on the primitives shown in Fig. \ref{Fig6Label} and their dimensions in Table \ref{tab:table1}, the threshold for using the predictor, $\gamma$, is set to be $0.03$. To optimize both the scatterers' shapes and the EMMS thickness, PSO is performed in each latent space separately and the best design is chosen by selecting the best global optimum. 

\begin{figure}[htbp]
	\setlength{\textfloatsep}{0.7\baselineskip plus 0.2\baselineskip minus 0.5\baselineskip}
	\centering 
  	\includegraphics[width=3.5in]{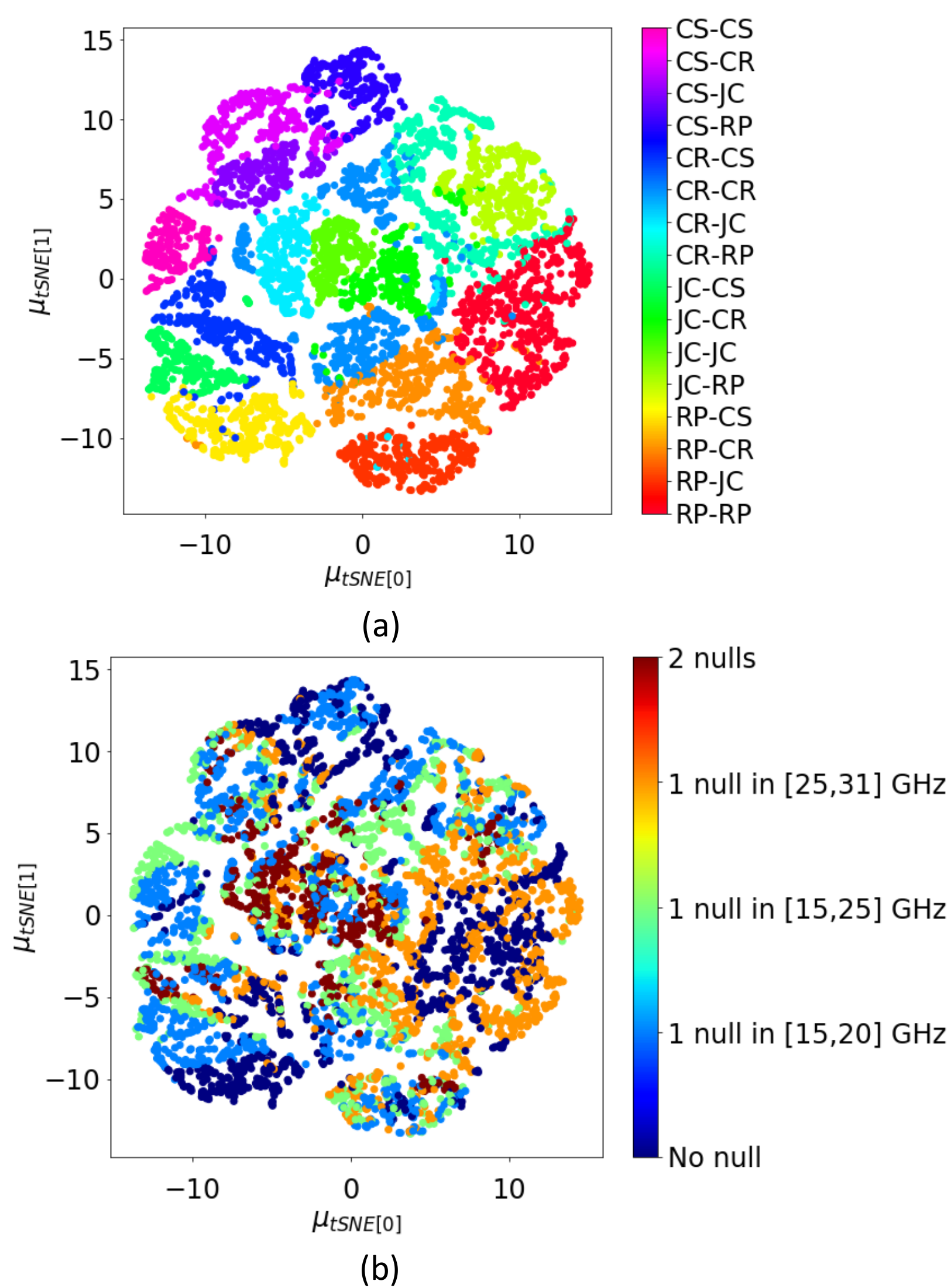}
  	\caption{The latent space of the $1.575$ mm dual-layer EMMSs in the training set in regards to (a) the top and bottom layer scatterers and (b) the number of nulls in the amplitude of the TE-transmission coefficient.}  	
  	\label{Fig7aLabel}
\end{figure}

The dimension $K$ of the latent space is tuned to be 8 for minimum reconstruction and prediction losses. For the purpose of visualization, this 8-dimensional latent space representing the dual-layer training set for substrate thickness of $1.575$ mm is converted into 2-dimensions, using t-distributed stochastic neighbor embedding (t-SNE) \cite{tSNE} and is shown in Fig. \ref{Fig7aLabel}. From Fig. \ref{Fig7aLabel} (a), it can be seen that the latent variables corresponding to EMMSs composed of the same type of scatterers, shown in Fig. \ref{Fig6Label}, are clustered together. Moreover, we can examine this latent space in relation to the scattering properties. While the amplitude and phase of the TE- and TM-transmission coefficients collectively impact the latent variables, let us look at the effect of the amplitude of the TE-transmission coefficients as an example. We categorize this property based on its number of nulls and their frequencies to 5 categories: no null, one null in the range  $[15-20]$ GHz, one null in the range $[20-25]$ GHz, one null in the range $[25-31]$ GHz, and two nulls. From Fig. \ref{Fig7aLabel} (b), we can see that there is also a correspondence between the amplitude of this coefficient and the latent variables.
    
\section{Design Examples} \label{section:V} \label{section:V}
\vspace{-1mm}
In this section, the potential of the proposed approach is demonstrated through various examples of dual- and three-layer EMMSs. Regardless of the application, the one-time generated training sets and trained ML models for dual- and three-layer EMMSs are used for the inverse design. 

These examples include asking for an EMMS to achieve a specific scattering response over the frequency range. However, in practice, the desired scattering properties are rarely described specifically over the frequency range. Instead, one might prefer to define the target scattering coefficients using minimum/maximum tolerable values in the transmission bands, stopbands, and/or phase difference between the responses of the EMMS for the two orthogonal polarizations. The proposed method can be used for this type of general synthesis as well. To do so, we use minimum, $[S]_ \textrm{min}$, and maximum, $[S]_ \textrm{max}$, amplitude and phase masks over frequencies to determine how close the latent variable's scattering properties are to the desired ones. Therefore, $|\Delta[S]|$ in \eqref{eq:6} is adjusted as follows: 
\begin{equation}\label{eq:7}
	\begin{array}{l}
|\Delta[S]| = \\
\quad \sum_{i=1}^{n}([S]_ \textrm{lv}(i)-[S]_ \textrm{min}(i))\times([S]_ \textrm{lv}(i)-[S]_ \textrm{max}(i)) \\
\qquad  + |([S]_ \textrm{lv}(i)-[S]_ \textrm{min}(i))\times([S]_ \textrm{lv}(i)-[S]_ \textrm{max}(i))| \\
	\end{array}
\end{equation}
It is worth noticing that if $[S]_ \textrm{lv}$ is between the indicated bounds by  $[S]_ \textrm{min}$ and $[S]_ \textrm{max}$, the $|\Delta[S]|$ expectedly becomes zero. Otherwise, it gets penalized by how much the $[S]_ \textrm{lv}$ is out of the bounds.
\begin{equation}\label{eq:7a} 
\begin{array}{l}
|\Delta[S](i)|\\
=
	\begin{cases}
	0 \hspace{5em}  \: [S]_ \textrm{min}(i) \le [S]_ \textrm{lv}(i) \le [S]_ \textrm{max}(i)  \\
	2 \times |([S]_ \textrm{lv}(i)-[S]_ \textrm{min}(i))\times([S]_ \textrm{lv}(i)-[S]_ \textrm{max}(i))| \\
	 \qquad \qquad \qquad \qquad  \qquad \qquad \qquad \qquad \qquad\text{elsewhere}
	\end{cases}
\end{array}
\end{equation} 

\vspace{-1mm} 
\subsection{Dual-Layer EMMS for Achieving a Specific Scattering Response}
This example includes asking for an EMMS that provides specific amplitude and phase of TE- and TM-transmission coefficients over the frequency range. We ensured that this set of coefficients are not achievable by any of the EMMSs used for the training of the ML models. Therefore, a new EMMS has to be generated to achieve the desired results. For this example, the optimization objectives are evaluated based on $|\Delta[S]$ in \eqref{eq:6}.

\begin{figure}[!ht]
	\centering 
  	\includegraphics[width=3.6in]{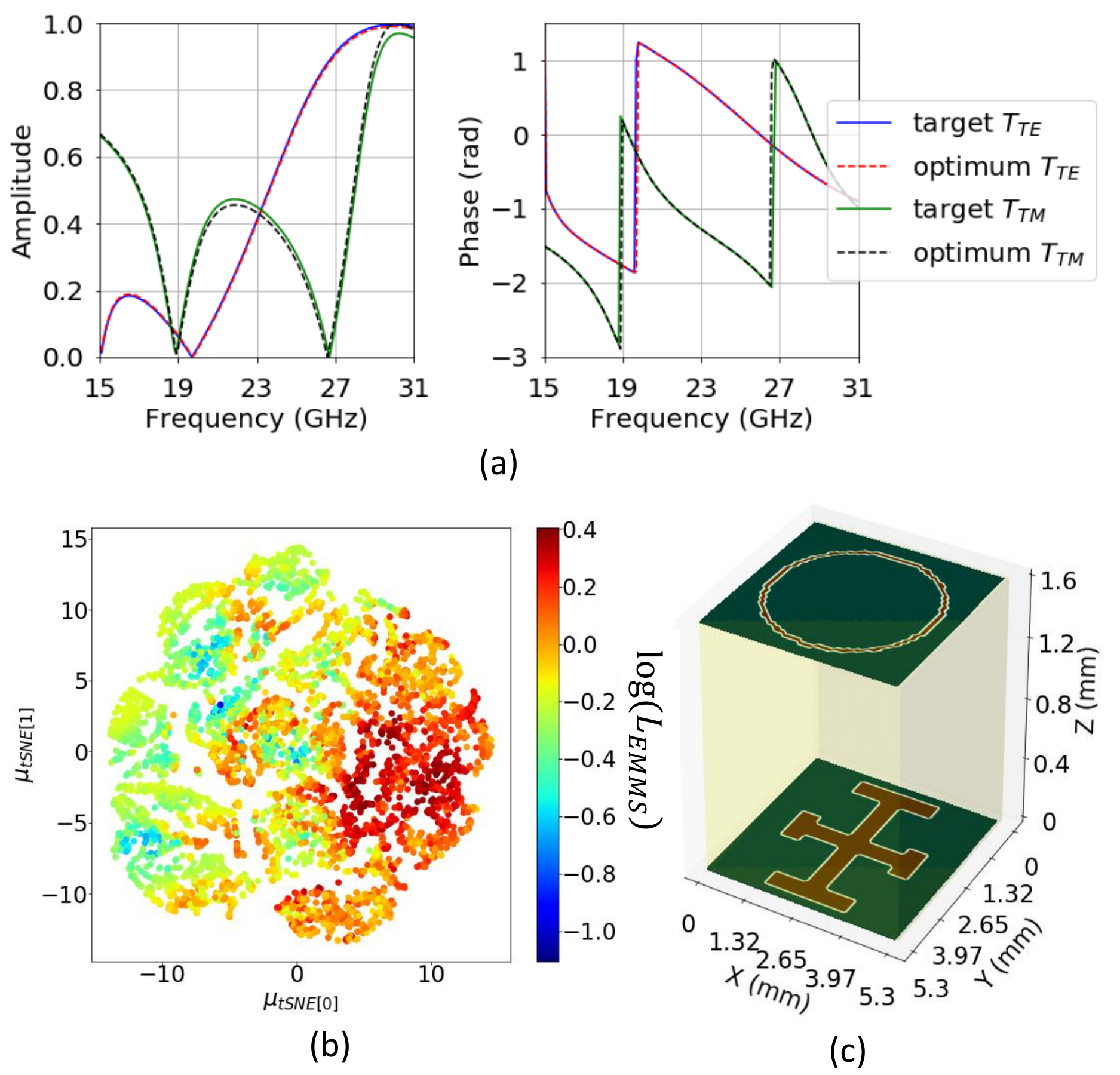}
  	\caption{(a) Target and optimized TE- and TM-mode transmission coefficients, (b) $\log_{10}(L_{EMMS})$ over the training t-SNE converted latent variables, and (c) optimum dual-layer EMMS.}
  	\label{Fig8Label}
\end{figure}

The amplitude and phase of the target and optimum TE and TM-mode transmission coefficients are shown in Fig. \ref{Fig8Label} (a), where an excellent match between the two is achieved. The logarithm of the difference between the transmission coefficients corresponding to the 2-dimensional t-SNE converted latent variables of the training data and the desired transmission coefficients, $\log_{10}(L_{EMMS})$, is plotted in Fig. \ref{Fig8Label} (b). This figure shows the expected smooth change of $L_{EMMS}$ over the regularized latent space. Moreover, it can be seen that the minimum value of the $\log_{10}(L_{EMMS})$ is $-1.10$, or minimum $L_{EMMS}=7.87 \times 10^{-2}$ over the training data, whereas the $L_{EMMS}$ corresponding to the optimum design is only $5.71 \times 10^{-4}$. This optimum response is provided by a newly generated dual-layer EMMS, shown in Fig. \ref{Fig8Label} (c). This dual-layer EMMS is the stacked Jerusalem cross and complete ring separated by $1.575$ mm of Rogers $5880$ dielectric. The dimensions of the Jerusalem cross are $l_{JC,x}=3.8$ mm and $l_{JC,y}=2.6$ mm and the rest of the parameters can be found in Table \ref{tab:table1}. The radius and width of the ring are $r_{CR}=2.2$ mm and $w_{CR}=0.1$ mm, respectively.
\vspace{3mm}

\subsection{Dual-Layer Dual-Band Frequency Selective Surface (FSS)}
A dual-layer EMMS is optimized as a TE-polarized dual-band frequency selective surface with a stopband between the two transmission bands. $\Delta[S]$ in \eqref{eq:7a} is only written for the magnitude of the TE-transmission coefficients while there is no restriction on the phase of the TE-transmission coefficient and the phase and magnitude of TM-transmission coefficient as 

\begin{equation}\label{eq:7aa} 
\begin{array}{l}
|\Delta[S](i)|\\
=
	\begin{cases}
	0 \hspace{5em}  \: |T_ \textrm{TE,min}(i)| \le |T_ \textrm{lv}(i)| \le |T_ \textrm{TE,max}(i)|  \\
	2 \times |(|T_ \textrm{lv}(i)|-|T_ \textrm{TE,min}(i)|)\times(|T_ \textrm{lv}(i)|-|T_ \textrm{TE,max}(i)|)| \\
	\qquad \qquad \qquad \qquad  \qquad \qquad \qquad \qquad \qquad  \text{elsewhere}
	\end{cases}
\end{array}
\end{equation} 

\begin{figure}[h!]
	\setlength\belowcaptionskip{-0.7\baselineskip}
	\centering 
  	\includegraphics[width=3.6in]{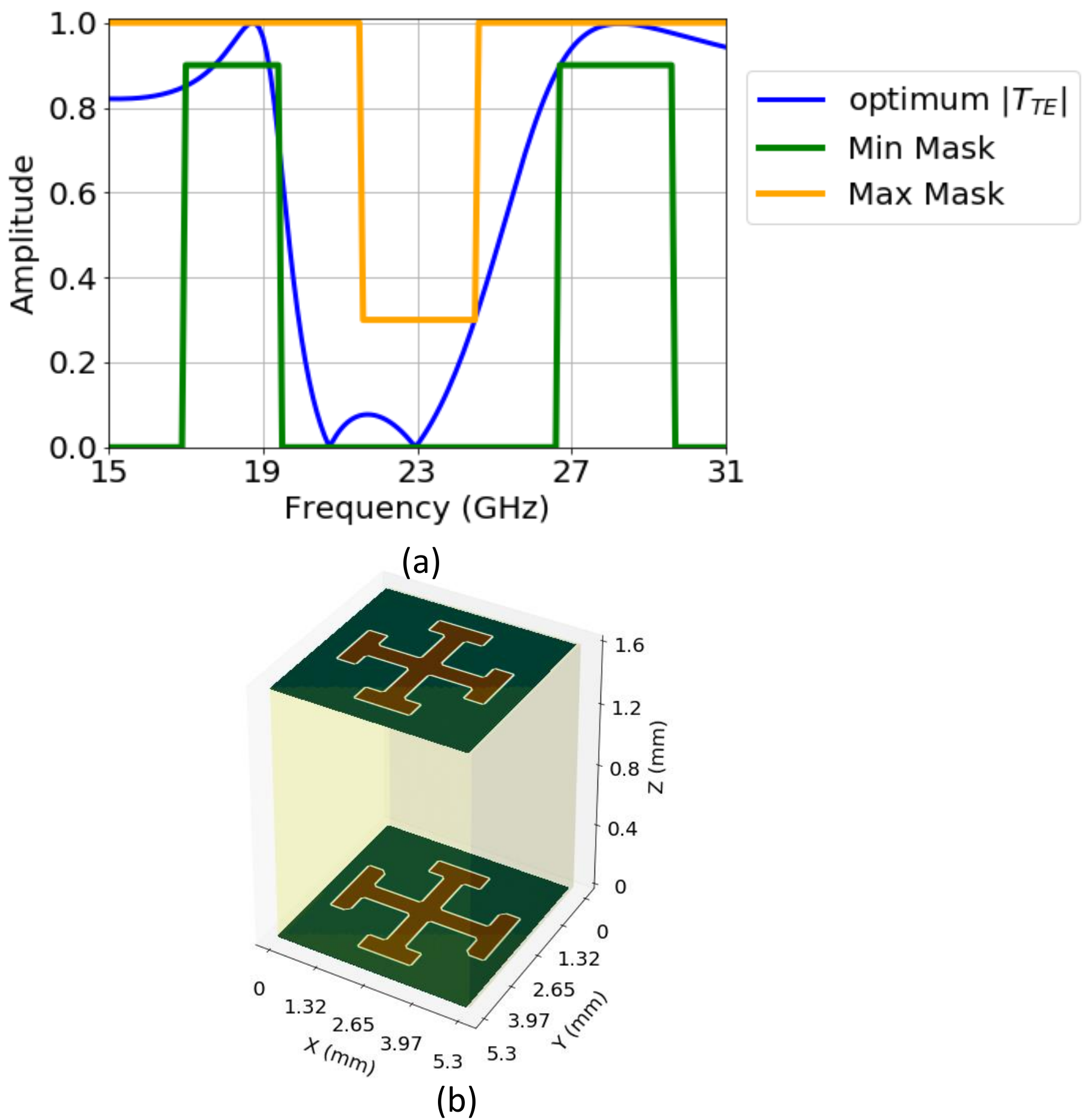}
  	\caption{Dual-layer dual-band FSS: (a) Minimum and maximum masks for magnitude of TE-mode transmission coefficient and the optimum magnitude of TE-transmission coefficient, and (b) optimum dual-layer EMMS.}
  	\label{Fig9Label}
\end{figure}

Fig. \ref{Fig9Label} (a) shows the defined minimum and maximum masks for a transmission coefficients whose magnitude is between $0.9$ and $1.0$ for $17.0-19.5$ GHz and $26.5-29.5$ GHz. Moreover, the stopband is in $21.5-24.5$ GHz where the amplitude is less than $0.3$. It can be seen that the optimum EMMS in Fig. \ref{Fig9Label} (b), meets the constraints with a slight mask violation in the lower band. The optimum design is a two Jerusalem crosses separated by a $1.575$ mm Rogers $5880$ dielectric. The dimensions of the top and bottom Jerusalem crosses are based on the physical parameters in Table \ref{tab:table1}, where $l_{JC,x}=l_{JC,y}=3.0$ and $l_{JC,x}=3.4$ mm and $l_{JC,y}=3.0$ mm, respectively. This optimum EMMS was generated to match the defined masks with much less error compared to the samples in the training data. It is worth noting that, in fact, meeting the specified constraints with less error is not possible with any dual-layer EMMS composed of a single scatterer on each layer due to the limited order of the response that can be achieved with only two layers.

\subsection{Three-Layer Wideband Linear-to-Circular Polarizer}
To design a linear-to-circular polarization converter (or simply, a polarizer), it is required that the magnitude of the TE- and TM-transmission coefficients would be ideally equal to $1.0$ over the desired frequency band, while maintaining a $\pm 90^\circ$ phase difference \cite{NaseriPol}. That way, if a $45^\circ$-tilted linear polarized incident wave excites the EMMS, the TE and TM components of the wave are transmitted fully with $\pm 90^\circ$ phase difference. Therefore, the transmitted wave would be a circularly-polarized wave. 

We design a polarizer that works from $20.5$ GHz to $30.5$ GHz. Based on the constraints explained above, we define two sets of masks for the magnitude and the phase difference between the two the transmission coefficients. The amplitude masks are defined to keep the magnitude of both transmission coefficient between $0.9$ and $1.0$ in the frequency band of interest. The phase masks are defined to keep the phase difference between $80^\circ$ and $100^\circ$. Three optimization objectives are defined for the amplitude of the TE- and TM- transmission coefficients and the phase difference between them, analogous to \eqref{eq:7aa}.

\begin{figure}[h!]
	\centering 
  	\includegraphics[width=3.6in]{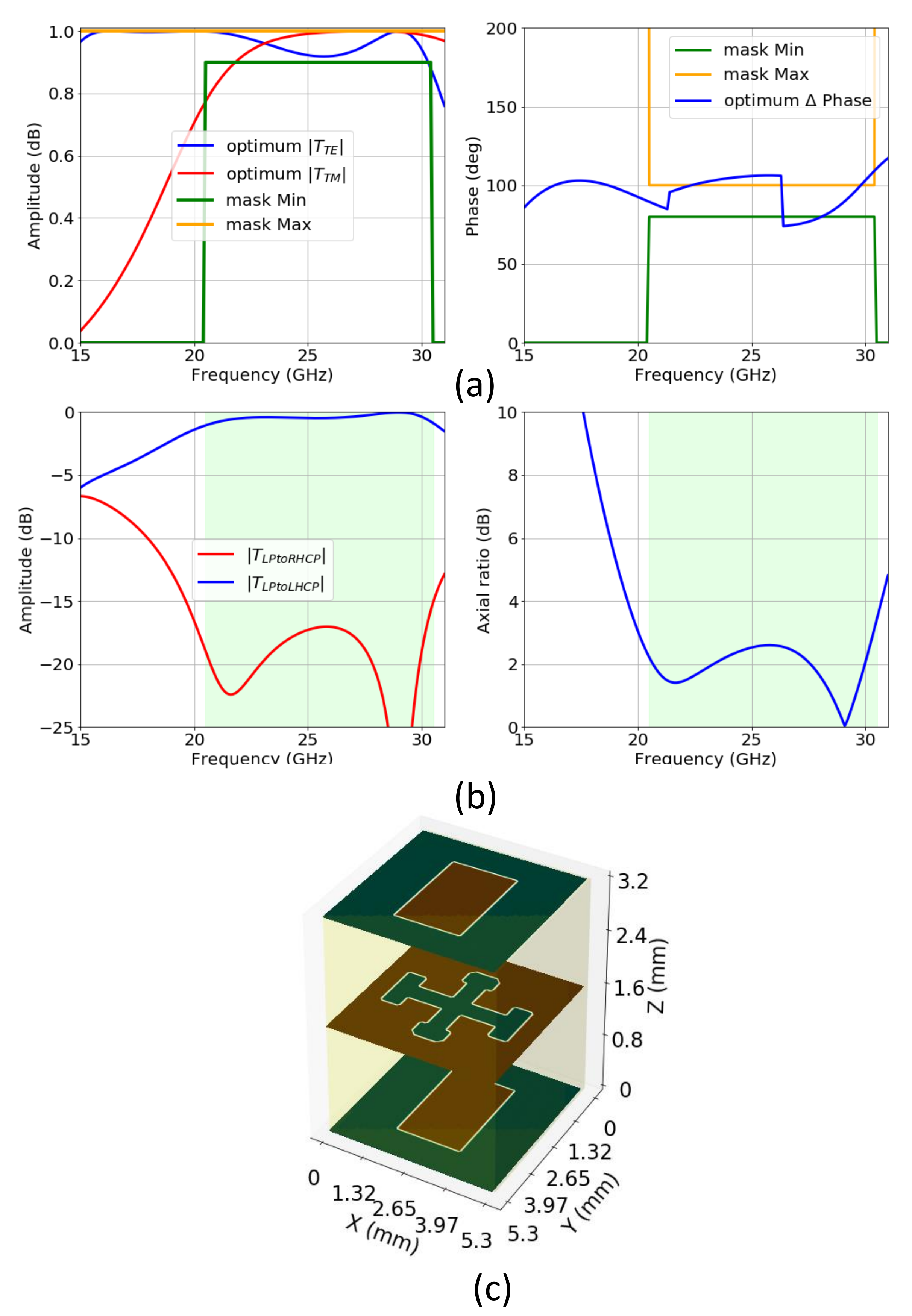}
  	\caption{Three-layer wideband linear-to-circular polarizer: (a) Minimum and maximum amplitude and phase difference masks for TE- and TM-transmission coefficients and the optimum transmission properties, (b) optimum linear-to-circular transmission coefficients and axial ratio, and (c) optimum three-layer EMMS.}
  	\label{Fig10Label}
\end{figure}

The masks and the optimum transmission coefficients are shown in Fig. \ref{Fig10Label} (a). Fig. \ref{Fig10Label} (b) shows that the linear-to-circular polarization conversion is successfully done by the optimized EMMS with less than $1$ dB od insertion loss and an axial ratio less than $3$ dB in the desired frequency band. The optimized three-layer EMMS composed of a rectangular patch with $l_{P,x}=3.2$ mm and $l_{P,y}=2.0$ mm on the top and bottom layers, and a complementary Jerusalem cross with the dimensions mentioned in Table \ref{tab:table1} with $l_{JC,x}=3.0$ mm and $l_{JC,y}=2.4$ mm on the middle layer. The scatterers are separated by two $1.575$ mm Rogers $5880$ dielectric substrates.

\subsection{Three-Layer Single Wideband FSS}

\begin{figure}[h!]
	\centering 
  	\includegraphics[width=3.6in]{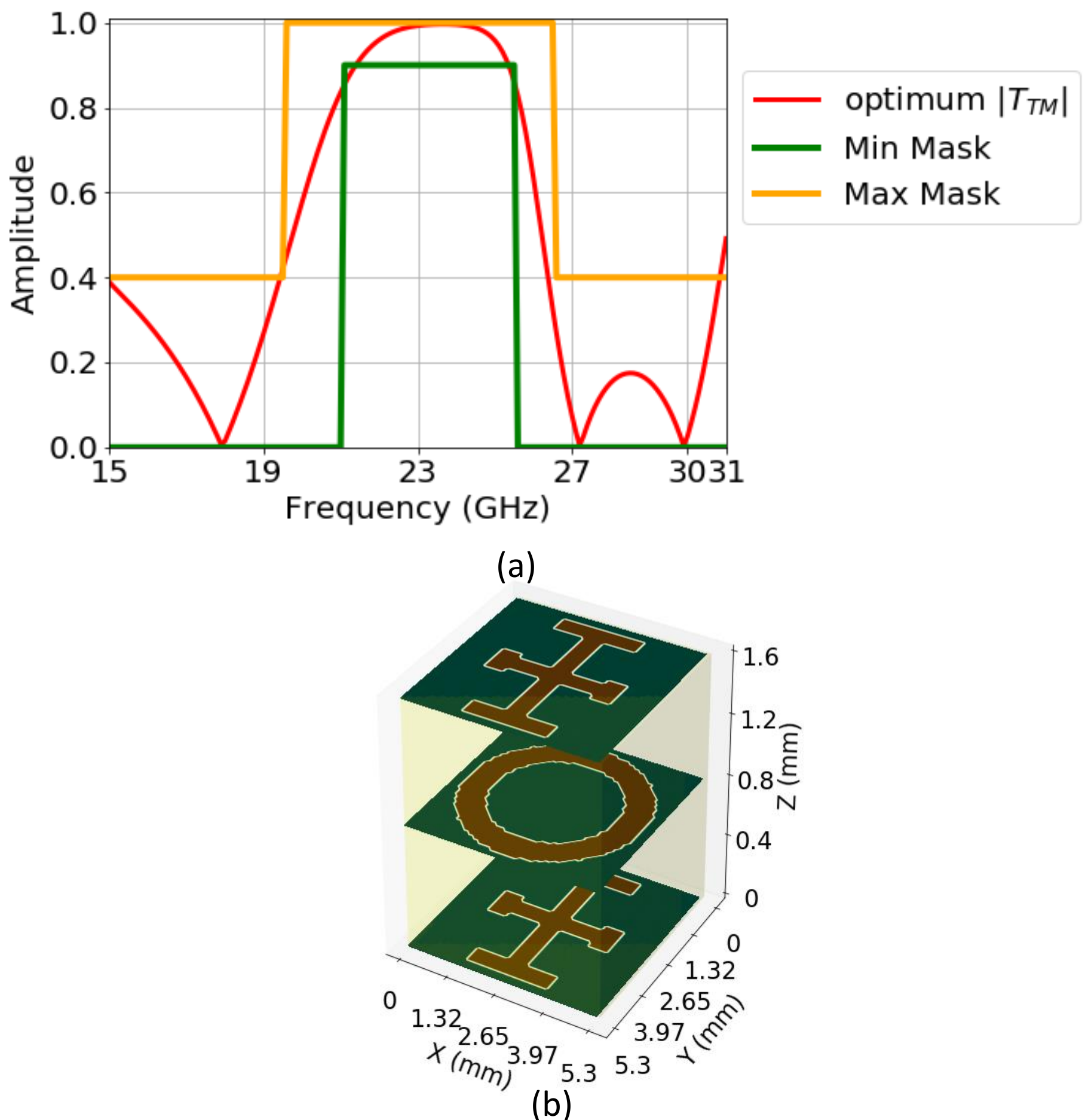}
  	\caption{Three-layer single wideband TM-FSS: (a) Minimum and maximum masks for magnitude of TM-mode transmission coefficient and the optimum magnitude of TM-transmission coefficient, and (b) optimum three-layer EMMS.}
  	\label{Fig11Label}
\end{figure}

In this example, a three-layer EMMS is optimized to filter the spectrum for the TM-mode incident waves in the $15.0-31.0$ frequency band such that TM-polarized waves between $21.0$ GHz and $25.5$ GHz are transmitted while those outside this interval get reflected. Similarly, two masks, shown in Fig. \ref{Fig11Label} (a), are defined to keep the TM-transmission coefficient amplitude above $0.9$ in the transmission band and below $0.4$ in the reflection bands. The optimization objective on the amplitude of the TM-transmission coefficient is evaluated similar to \eqref{eq:7aa}.  The amplitude of the optimum TM-transmission coefficient and the optimum design are shown in Fig. \ref{Fig11Label} (a) and (b), respectively. 

The optimum EMMS is a $1.574$ mm thick structure composed of two Jerusalem crosses on the top and bottom layers and a ring in the middle layer that is embedded between two $0.787$ mm Rogers 5880 substrates. The dimensions of the Jerusalem cross are listed in Table \ref{tab:table1}, where $l_{JC,x}=2.4$ mm and $l_{JC,y}=3.8$. The radius and width of the ring are $r_{CR}=2.4$ mm and $w_{CR}=0.7$ mm.   This EMMS was generated to match the constraints with minimum error. It is worth noting that the third resonance at $30$ GHz is important to keep the transmission amplitude below $0.4$ in the higher band. This resonance is caused by the higher-order coupling between the scaterrers in the EMMS. Therefore, besides the scatterers' shapes and dimensions, the thin substrate and higher-order coupling had to be optimized to achieve satisfactory results, demonstrating the utility of the proposed approach.      

\section{Exploration in the Design Space}\label{section:VI}
\begin{figure}[h!]
	\centering 
  	\includegraphics[width=3.6in]{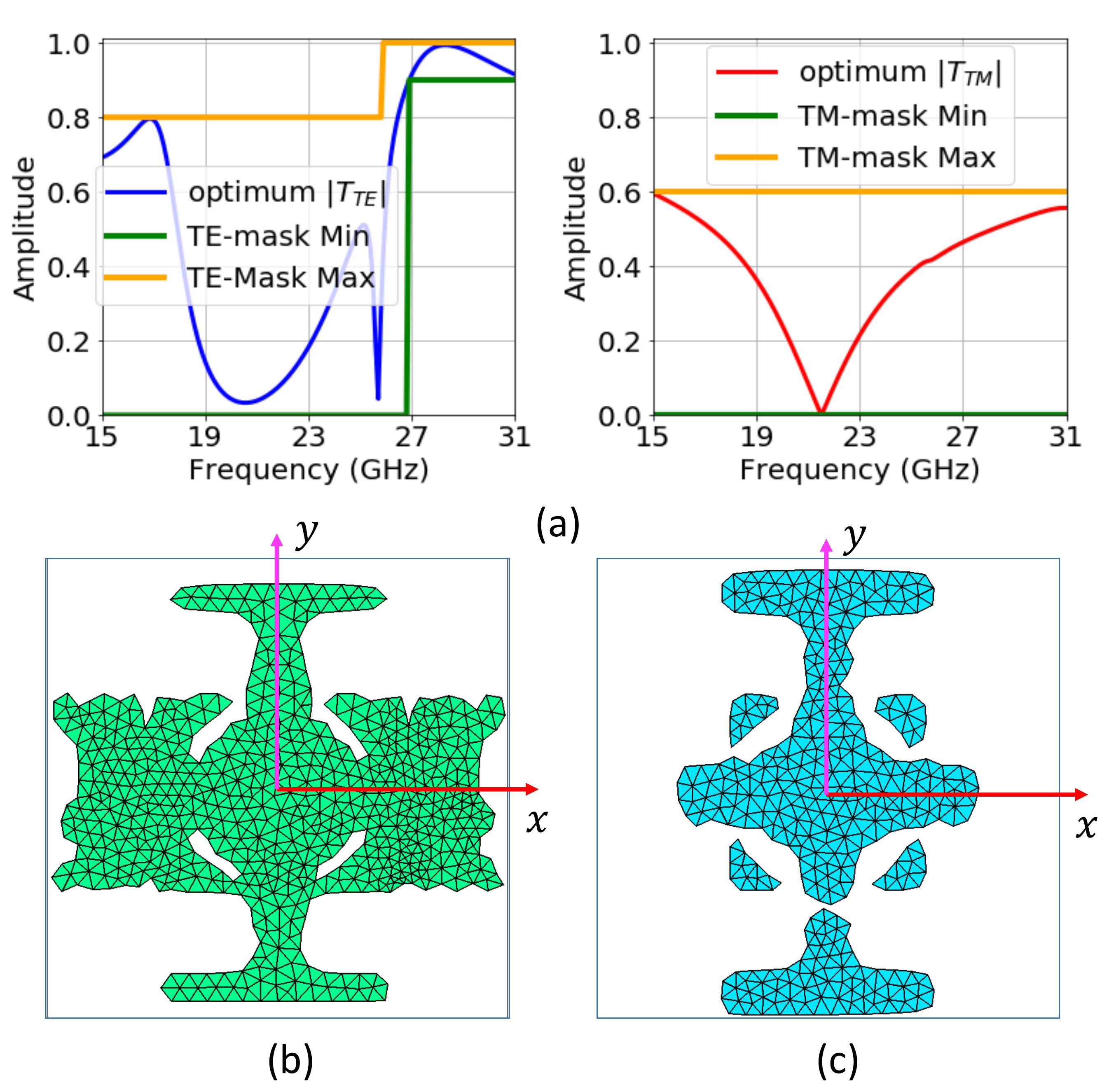}
  	\caption{(a) Minimum and maximum masks for the amplitude of the TE- and TM-transmission coefficients, (b) the top, and (c) the bottom layers of the optimum dual-layer EMMS with $1.575$ mm dielectric separation.}
  	\label{Fig12Label}
\end{figure}

Using the primitives shown in Fig. \ref{Fig6Label} (a)-(d) with the dimensions specified in Table \ref{tab:table1}, about $178,000$ dual-layer EMMS samples can be generated that provide a wide variety of TE- and TM-responses. However, here, we deliberately define two arbitrary sets of amplitude masks for the desired TE- and TM- response, shown in Fig. \ref{Fig12Label} (a), to push the algorithm to a more ``generative mode'' where new scatterer shapes have to be generated to meet all the indicated requirements. The purpose of this example is to show that if the desired properties are not met by stacking the known scatterers shown in Fig. \ref{Fig6Label}, new shapes of scatterers will be generated by the proposed approach to meet the requirements. As mentioned before, in this case, the images of the scatterers of the optimum design are converted to meshes using the RWG basis functions and simulated. The GSMs of both scatterers are cascaded to obtain the scattering properties of the dual-layer structure. The meshed structures of the scatterers on the top and bottom layers of the optimum EMMS are shown in Fig. \ref{Fig12Label} (b) and (c), respectively. These scatterers are separated by a $1.575$ mm Roger $5880$ dielectric substrate. 

The scatterer on the top layer of this optimum EMMS, shown in Fig. \ref{Fig12Label} (b), can be considered as the interpolation of a rectangular patch and a Jerusalem cross. The scatterer on the bottom layer, shown in Fig. \ref{Fig12Label} (c), appears to be an interpolation of a Jerusalem cross and the circular slot. That is why the latent variable corresponding to the combination of the two scatterers lies in one of the gaps of the latent space where it is not covered by the samples of the training set. Therefore, the continuous representation of the EMMSs in the latent space makes it possible to easily interpolate different structures and obtain brand-new scattering properties. 

\section{Conclusion} \label{section:Conclusion}
A machine learning-based approach has been proposed to solve the \emph{inverse problem} of designing multilayer metasurfaces based on the desired scattering properties. Using a generative ML model based on a variational autoencoder, the structures and the scattering properties of the multilayer EMMSs in the training set are converted to a low-dimensional continuous latent space. In this latent space, the particle swarm optimization has been performed to find the optimum latent variable and consequently the physical design of the optimum EMMS. This approach exploits the information learned from the training set and explores the design space by interpolating the structures and the properties of the training structures in the latent space to propose new EMMSs that meet the desired requirements. 

The EMMS optimization objectives are evaluated using efficient method including using ML surrogate models when they are known to provide sufficient accuracy, fast cascading the GSMs of the known scatterers, and full-wave simulations for brand-new generated structures. Using this method, we eliminate the need for very large amounts of training data that are required for thin EMMSs where interlayer coupling and structural modifications can change the scattering properties significantly. Therefore, while providing reliable assessment of the structure, we expedite the evaluation process as much as possible with the ML surrogate models and cascading process, and perform full-wave simulations only if the proposed EMMS is likely to meet the requirements. Hence, we were able not only to remove brute-forcing the combinations of the known scatterers, but also to explore the design space by generating new structures. Different examples of multiobjective optimizations to achieve specific dispersive TE- and TM-responses and application-based criteria such as dual-band frequency selective surface, wideband liner-to-circular polarization, and wideband FSS by different EMMSs have been demonstrated. Using individual latent spaces to represent dual-and three-layer EMMSs with different dielectric thicknesses, we were able to find the global optimum by adjusting both the scaterrers and the dielectric thicknesses. 

The proposed approach here can be extended to inverse design of metasurfaces with more than two choices of scatterers including three-layer bianistropic surfaces and surfaces with more than three layers. Moreover, other macroscopic properties of metasufaces such as surface admittance/impedance and susceptibilities can be defined as desired targets based on the application at hand.


%

\appendices
\section{Neural Network Details} \label{appendix: B}
The representation of the EMMSs and their transmission coefficients can be turned to a low-dimensional latent space. For that, we use a jointly trained VAE and predictor, shown in Fig. \ref{Fig4Label}.  The two $52 \times 52$-images of the resonators of the EMMS are flattened and concatenated together to form a $1 \times 5704$-vector. This vector, denoted by $x$, has $0$ (no metal) or $1$ (metal) values for its components and is used as the input of the encoder. All the neural networks are implemented with multilayer perceptrons (MLPs). In the following paragraphs, we outline their number of hidden layers, neurons and activation functions. The encoder has $8$ hidden layers with $\{2048, 2048, 1024, 512, 512, 256, 128, 64\}$ neurons and rectified linear unit function (ReLU) activation function. The $8$th-hidden layer is connected to two $8$-dimensional hidden layers as the mean and variance layers, shown in Fig. \ref{Fig4Label}. The dimension of the latent space is optimized to be $8$ for acceptable reconstruction and prediction losses. The decoder has $8$ hidden layers with $\{64, 128, 256, 512, 512, 1024, 2048, 2048\}$ neurons and the ReLU activation function. The output layer of the decoder, denoted by $\hat{x}$, has $5704$ neurons with the sigmoid activation function to reconstruct the input. It is worth noting that the sigmoid activation function is used to create values between $0$ and $1$. 

Frequency points between $15$ and $31$ GHz are normalized to values between $0$ and $1$ for better training and used along the $8$-dimensional latent variables as inputs of the predictor.  Here, for simpler implementation, we use four predictors to output the amplitude and phase of the TE- and TM- transmission coefficients, named \emph{magPredictor} and \emph{phasePredictor}, respectively. The \emph{magPredictor}s have $7$ hidden layers with $\{500, 1000, 2000, 1000, 500, 200, 100\}$ neurons and the ReLU activation function. Since the amplitude of transmission coefficient has a value between $0$ and $1$ at each frequency, the sigmoid activation function is used for the one-dimensional output layer. The \emph{phasePredictor}s have $8$ hidden layers with $\{100, 200, 500, 1000, 1000, 500, 200, 100\}$ neurons and the ReLU activation function that outputs the normalized transmission phase with the sigmoid activation function.

During the training process, the mean and variance hidden layers at the end of the encoder in addition to the the weights of the MLPs, are optimized using the Adam optimizer \cite{Adam} with the learning rate of $l_r=0.0005$ to minimize the loss function in \eqref{eq:3}. Since the input is images with $0$ and $1$ values, the reconstruction loss, $L_{recons}$ be calculated using \eqref{eq:4}. $\alpha$ and $\beta$ are tuned to $10$ and $20$, respectively. It is worth mentioning that we used gradually increasing the batch size to reduce the training loss further. 


\section*{Acknowledgment}

The authors would like to thank Zhengzheng Wang for his help with developing techniques to accelerate our in-house MoM code to expedite the simulations required for this project.

\ifCLASSOPTIONcaptionsoff
  \newpage
\fi

\end{document}